\documentclass[11pt]{article}
\usepackage{amsfonts,amssymb,epsfig}
\usepackage[latin1]{inputenc}
\usepackage[english]{babel}
\usepackage{color}
\usepackage{hyperref}
\newtheorem{thm}{Th\'eor\`eme}[section]
\newtheorem{cor}[thm]{Corollaire}
\newtheorem{lem}[thm]{Lemme}
\newtheorem{pro}[thm]{Proposition}
\newtheorem{dfn}[thm]{D\'efinition}

\newtheorem{expl}[thm]{Exemple}

\def\dessous#1\sous#2{\mathrel{\mathop{\kern0pt#2}\limits_{#1}}}

\newcommand{\beq}{\begin{eqnarray}}
\newcommand{\eeq}{\end{eqnarray}}
\newcommand{\bpro}{\begin{pro}}
\newcommand{\epro}{\end{pro}}
\newcommand{\blem}{\begin{lem}}
\newcommand{\elem}{\end{lem}}
\newcommand{\bdfn}{\begin{dfn}}
\newcommand{\edfn}{\end{dfn}}
\newcommand{\bcor}{\begin{cor}}
\newcommand{\ecor}{\end{cor}}
\newcommand{\bthm}{\begin{thm}}
\newcommand{\ethm}{\end{thm}}
\newcommand{\bex}{\begin{expl}}
\newcommand{\eex}{\end{expl}}
\newcommand{\brmq}{\begin{rmq}}
\newcommand{\ermq}{\end{rmq}}
\newcommand{\benum}{\begin{enumerate}}
\newcommand{\eenum}{\end{enumerate}}
\newcommand{\bitem}{\begin{itemize}}
\newcommand{\eitem
}{\end{itemize}}
\begin{document}

   \begin{center}

{\Large {Effects of critical fluctuations and dimensionality on the jump in specific heat at the superconducting transition temperature: Application to $\mathrm{YBa}_2\mathrm{Cu}_3\mathrm{O}_{7 - \delta}$, $\mathrm{Bi}_{2}\mathrm{Sr}_{2}\mathrm{CaCu}_{2}\mathrm{O}_{8+ \delta}$ and $\mathrm{KOs}_2\mathrm{O}_6$ compounds}}

 \vspace{1cm}

 \bf R. M. Keumo Tsiaze$^{a, b}$, A. V. Wirngo$^{b}$, S. E. Mkam Tchouobiap$^{b, c}$, A. J. Fotue$^{a, d}$,
 
  E. Balo\"{\i}tcha$^{b}$ and M. N. Hounkonnou$^{b}$
 
\vspace{1cm}

{\em  $^{a}$Laboratory of Mechanics, Materials and Structures, Department of Physics, Faculty of Science, University of Yaound\'e I - P.O. Box 812, Yaound\'e, Cameroon.}\\
{\em  $^{b}$International Chair in Mathematical Physics and Applications (ICMPA-UNESCO Chair), University of Abomey-Calavi, 072 B.P. 50, Cotonou, Republic of Benin.}\\
{\em $^{c}$ Laboratory of Research on Advanced Materials and Nonlinear Science (LaRAMaNS),Department of Physics, Faculty of Science, University of Buea,  P.O. Box 63, Buea, Cameroon.}\\
{\em $^{d}$ Mesoscopic and Multilayer Structures Laboratory, Department of Physics, Faculty of Science, University of Dschang - P.O.Box 479, Dschang, Cameroun.}\\

{\em E-mail: {\tt keumoroger@yahoo.fr, hounkonnou@yahoo.fr
}}

\begin{abstract}
\large{We report on a study of the superconducting order parameter thermodynamic fluctuations in
$\mathrm{YBa}_2\mathrm{Cu}_3\mathrm{O}_{7 - \delta}$, $\mathrm{Bi}_{2}\mathrm{Sr}_{2}\mathrm{CaCu}_{2}\mathrm{O}_{8+ \delta}$ and  $\mathrm{KOs}_2\mathrm{O}_6$ compounds. A non-perturbative technique within the framework of the renormalized Gaussian approach is proposed. The essential features are reported (analytically and numerically) through Ginzburg-Landau (GL) model-based calculations which take into account both the dimension and the microscopic parameters of the system. By presenting a self-consistent approach (SCA) improvement on the GL theory, a technique for obtaining corrections to the asymptotic critical behavior in terms  of non universal parameters is developed. Therefore, corrections to the specific heat and the critical transition temperature for one-, two- and three-dimensional samples are found taking into account the fact that fluctuations occur at all length scales as the critical point of a system is approached. The GL model in the free-field approximation and the 3D-XY model are suitable for describing the weak and strong fluctuation regimes respectively. However, with a modified quadratic coefficient, the renormalized GL model is able to explain certain experimental observations including the specific heat of complicated systems, such as the cuprate superconductors and the $\beta$-pyrochlore oxides. It is clearly shown that the enhancement, suppression or rounding of the specific heat jump of high-$T_c$ cuprate superconductors at the transition are indicative of the order parameter thermodynamic fluctuations according to the dimension and the nature of interactions.}
\end{abstract}
\end{center}
\large{ \baselineskip 8mm \noindent \textbf{Keywords:}
\textit{Superconducting phase transitions, cuprate superconductors, low-dimensional systems, critical fluctuations, Ginzburg-Landau theory, self-consistent approach}}.\\


\section{\bf {Introduction}}
\noindent

A survey of phase transitions and critical phenomena in cuprate superconductors was given by Schneider
et al \cite{schneider}. Close to the critical point, the order parameter fluctuates on all length scales and those fluctuations smear out the microscopic details of the interactions in the system. Scalapino, Sears and Ferrell \cite{scalapino} showed that the failure of the mean-field theory (MFT) to provide an accurate description of critical behavior and to take into account both the microscopic details of the interactions and the dimensionality of physical systems is due to an improper treatment of the aforementioned fluctuations \cite{salamon}. Indeed, if MFT satisfactorily accounts for some key features of the transition mechanism at the instability point and predicts a finite mean-field critical temperature $T_c$ \cite{landau, stanley}, there are some important critical phenomena which cannot be efficiently accounted for with the standard MFT. The Mermin-Wagner-Hohenberg theorem \cite{mermin, hohenberg} for instance stipulates that a  broken  continuous symmetry prevents long-range order in one or two dimensions, which means that the thermodynamic parameters of a system are dependent to a large extent on the geometry and the dimension of space.  This essential point can be well appreciated from the fact that due to intrinsic fluctuations, a model with dipolar interactions alone exhibits no phase transition down to the lowest temperature as reported by Onsager \cite{onsager}.

 Intensive studies have been carried out in order to elucidate the relevance of critical phenomena, especially critical fluctuation effects in materials and physical systems in general, and further, to study the critical fluctuation dynamics and their influence on the physical properties of real materials. The studies are performed based on several approaches, which include the self-consistent MFT, self-consistent phonon approximation, low temperature approximation, standard Gaussian approach (SGA), dynamical MFT, dynamical self-consistent theory, etc. However, some of these attempts seem to give divergent and ambiguous conclusions. A common problem remains the difficulty to assign the parameters of a model to their characteristic quantities such as temperatures in a universal fashion \cite{cowley}. Accordingly, the idea behind the perturbation theory is to consider an exactly solvable model, either the gaussian or the mean-field model, and to introduce a correction to it by adding through a perturbation expansion, term(s) which is (are) not taken into account in the exactly solvable model. But, the coupling constant with the $\phi^4$ model of perturbation is not necessarily small, and as such the convergence of the perturbation expansion cannot be ensured.  Thus, some more effective approaches to the calculation are needed.
   
  Despite the significant experimental and theoretical effort to understand superconductivity and its nature in materials, there are still a number of open issues that have not been well addressed. From the theoretical point of view especially,  a number of properties have not been investigated. The temperature dependent electronic specific heat $C_{\textrm{el}}(T)$ measured through the superconducting transition is one such property. The study of $C_{\textrm{el}}(T)$ often has complications caused by the need for a careful subtraction of the normal state contributions, the size of the jump in $C_{\textrm{el}}$ at the  superconducting transition is known to depend on the microscopic details of the superconducting state \cite{salamon, bok}.
  
 More precisely, the relevance of superconducting fluctuations with short wavelength implies that the conventional MF spectrum of the GL functional is no longer a valid approximation for superconducting cuprates because thermal fluctuations of the order parameter near the critical point are large due to the short coherence length, which follows directly from the high $T_c$ through Heisenberg's uncertainty principle \cite{gauzzi}.  For example, depending on the nature of fluctuations, the dimension and the lattice type, one observes in certain cuprate superconductors, a decrease in $T_c$ by a factor, $\tau$ (where $1 < \tau < 2$) \cite{loram}, an absence of the heat capacity jump at the transition as in $\mathrm{Bi}_{2}\mathrm{Sr}_{2}\mathrm{CaCu}_{2}\mathrm{O}_{8+ \delta}$ ($\mathrm{BSCCO}$) \cite{meingast, deutscher}, or an enhancement of this specific heat jump near the critical point, which in the latter may be evidence of another type of phase transition observed in a new class of transition metal oxide superconductors, namely the so-called $\beta$-pyrochlore oxides \cite{yonezawa, yonezawa2, zenji}. It is thus abnormal to believe that non universal features of critical phase transition can be described by a simple GL functional ($F_{GL}$). Instead, in order to find explanations to all these observations, it is important to obtain the explicit form of the effective $F_{GL}$. Phase transitions in physical systems can be described by an appropriate choice of the coefficients in the expansion of $F_{GL}$, provided that a sufficiently large number of terms (quartic and higher-order terms) and microscopic parameters of the system are taken into account. This observation is valid for both weak and strong fluctuations. 
    
 In this work, we show that $F_{GL}$ can be improved by the confinement of critical fluctuations (at the Gaussian level) by bringing to light a non universal parameter $\upsilon (d, T)$. We propose a different way of further clarifying those long-standing issues, at least for reduced temperature in the crossover region where it is known that MFT fails to explain the critical behavior. While analyzing the fluctuation effects on the electronic specific heat within the GL theory, we take into account the mode-mode interactions with different types of couplings. The value of $\upsilon (d, T)$ takes into account critical fluctuations and the dimension dependence of the quadratic coefficient of $F_{GL}$, and will be used as a criterion of adiabaticity. In this context, $T_c$ appears as a characteristic scale of temperature related to thermal fluctuations and finite-size effects rather than the transition temperature \cite{roger, dikande}. As raised in Ref. \cite{roger}, another technical problem consists in carrying out statistical averages over thermal fluctuations on all size scales.  Therefore, the method introduced is similar to that described by Wilson \cite{wilsonkg}, and consists in integrating out the fluctuations sequentially, starting with fluctuations on a microscopic scale and then moving to successively larger scales until fluctuations on all scales have been averaged out. In the case of superconductors, the passage from individual Cooper pairs to mesoscopic and macroscopic superconducting fields necessarily involves a certain smooth spatial averaging procedure, the main idea being to show that the SCA improved GL theory is able to explain certain experimental observations including the specific heat of complicated systems, such as the cuprate superconductors and the $\beta$-pyrochlore oxides.
    
	We emphasize that in this paper, we only refer to the zero field scaling regime which is intermediate between MFT and the asymptotic scaling regime inaccessibly close to the critical point where fluctuations of  the electromagnetic field must also be taken into account \cite{kamal}. The paper is organized as follows. In Sec. II, we introduce the model for the effects of order parameter fluctuations (OPF) on the transition temperature for 1D, 2D, 3D and $n$D material samples, respectively.  In Sec. III, we discuss the effects of the OPF on the specific heat (jump) and present some applications of the work by comparing our results to well-known results for homogeneously disordered systems, amongst which are $\mathrm{YBa}_2\mathrm{Cu}_3\mathrm{O}_{7 - \delta}$ ($\mathrm{YBCO}$), $\mathrm{BSCCO}$ and $\beta$-pyrochlore oxide compounds such as $\mathrm{KOs}_2\mathrm{O}_6$. The specific heat of these compounds can be qualitatively and quantitatively explained using the theoretical framework established here. Earlier results concerning fluctuation effects in high-$T_c$ superconductors (HTS) were reviewed by Salamon \cite{salamon}. Here, we discuss the results concerning specific heat jump expressions in cuprates in the absence of magnetic fields, which allow the elucidation of the nature of the phase transition in HTS. We also show that the rounding or the enhancement of the specific heat jump of high-$T_c$ cuprate superconductors is indicative of OPF effects. Finally, the conclusion is drawn in Sec. IV.

%%%%%%%%%%%%%%%%%%%%%%%%%%%%%%%%%%%%%%%%%%%%%%%%%%%%%%%%%%%%%%%%%%%%%%%%%%%%%%%%%%%%%%%%%%%%
\section{Model Formulation and Renormalized Gaussian approach}
%%%%%%%%%%%%%%%%%%%%%%%%%%%%%%%%%%%%%%%%%%%%%%%%%%%%%%%%%%%%%%%%%%%%%%%%%%%%%%%%%%%%%%%%%%%%%%
%
\noindent
The phenomenological theory of superconductivity is due to Ginzburg and Landau \cite{schneider} who realized that for a superconducting phase, the natural order parameter is the condensate wave function ${\Psi}(\textbf{r})$. The order parameter is a complex scalar, which in terms of its magnitude $\vert{\Psi}(\textbf{r})\vert$ and phase $\varphi(\textbf{r})$ \cite{anatoly} or its real $\Psi_R(\textbf{r})$ and imaginary $\Psi_I(\textbf{r})$ parts can be written as
\begin{equation}
\Psi(\textbf{r}) = \vert{\Psi}(\textbf{r})\vert\exp(i\varphi(\textbf{r})) = \Psi_R(\textbf{r}) + i\Psi_I(\textbf{r}).
\end{equation}
An isotropic superconductor corresponds to the macroscopic wavefunction of the pairs with charge 2$e$ and effective mass $m$. Hence $\vert{\Psi}(\textbf{r})\vert^2$ is the probability of finding a pair at the position vector $\textbf{r}$. Some superconductors, in particular the cuprates, exhibit a pronounced anisotropy in their superconducting and normal state properties. Indeed cuprates can be viewed as a stack of superconducting layers (parallel to the crystallographic $ab$-plane, i.e. the $xy$-plane) with relatively weak interlayer interactions in the $c$- or $z$-direction. The phenomenological properties of layered superconductors are usually well described by anisotropic effective models such as the Lawrence-Doniach model \cite{doniach}. An alternative way of introducing
the effect of the layering is to consider a GL theory having anisotropic masses \cite{tinkham}. Near  the  critical  temperature, the superconducting wave function $\Psi$ is small, so that the free-energy difference $F_{GL}[\Psi]$ between the normal and superconducting states can be expanded  in powers of $\vert{\Psi}(\textbf{r})\vert^2$ and one way of treating this anisotropy is to replace the effective mass, $m$, by a diagonal effective mass tensor, so that: 
\begin{equation}
 F_{GL}[\Psi] = \int{d^dr} \Big[\frac{\hbar}{2m_{\parallel}}\big(\vert\nabla_x{\Psi}\vert^{2} + \vert\nabla_y{\Psi}\vert^{2} \big) + \frac{\hbar}{2m_{\perp}}\vert\nabla_z{\Psi}\vert^{2}  + a(T)\vert{\Psi}\vert{^2}  + \frac{b_0}{2}\vert{\Psi}\vert{^4} + \cdots \Big].
\end{equation}
The  phenomenological  coefficients  $a(T) = a_0T_c\epsilon$ where $\epsilon = \frac{T - T_c}{T_c}$ is the mean-field reduced temperature in the absence of fluctuations and  $b_0 > 0$  appearing in Eq. (2) can be derived in general from experiments or from the microscopic functional: the masses $m_{\parallel, \perp}$ are those of the Cooper pairs. The gradient term takes into account possible spatial inhomogeneities in the system and corresponds to the continuum limit \cite{zinn-Justin, ginzburg}. In this approximation, the free-energy functional describes the second-order phase transition from the normal to the superconducting state at $T_c$ with the MF critical indices.

The cuprates exhibit a pronounced uniaxial anisotropy characterized by
\begin{equation}
m_z \equiv m_{\perp}, \hspace{0.5cm} m_x \approx m_y \equiv m_{\parallel}, \hspace{0.5cm} m_{\perp} \gg  m_{\parallel} .
\end{equation}
 At the mean-field level, this anisotropy does not modify the specific heat jump and the temperature dependence of the order parameter. It affects, however, the characteristic length scales. Therefore, using the Gaussian approximation where $F_{GL}$ is truncated to second order, the specific heat (obtained by making a transition to the continuum domain) can be factorized as follows \cite{roger}:
\begin{eqnarray}
&&C_{GL} =  \left\{  \begin{array}{ll}\vspace{0,5cm}
\frac{1}{2}\eta_d{(\xi_0^+)^{-d}}\Big[\frac{T}{T_c}\Big]^2\epsilon^{-\alpha}, & \textrm{for}\phantom{...} T > T_c \\
\Delta C_0 +   \eta_d{(\xi_0^-)^{-d}} \Big[\frac{T}{T_c}\Big]^2\vert\epsilon\vert^{-\alpha'}, & \textrm{for} \phantom{...} T < T_c\\
  \end{array}\right.
\end{eqnarray}
where  
\begin{equation}
\eta_d
=\frac{k_B}{\Gamma(d/2)}\frac{L^{d}}{(2\pi)^{d/2}}\int_0^{+\infty} \frac{x^{d-1} dx}{(1+x^2)^2},
\end{equation}
is an integral quantity implied by the Gaussian approximation which depends on the dimensionality, and
\begin{equation}
\Delta C_0 = \frac{a^2_0 T_c}{b_0},
\end{equation}
is the jump of the resulting mean-field specific heat as $T$ decreases below $T_c$. $\xi_0^{\pm}$ are not universal quantities and $\xi_0^{+}$ differs from $\xi_0^{-}$ only by the factor $\sqrt{2}$. The dominant behavior of $C_{GL}$ close to $T_c$ is obtained through the power law \cite{ma} $C_{GL} \sim {\vert\epsilon\vert}^{d\nu-2}$, which yields the (Gaussian) critical exponents $\alpha = \alpha' = 2-d\nu$ with $\nu = 1/2$. Generally, the contribution of Gaussian fluctuations to the specific heat is given by $C_{GL} = C^{\pm}\epsilon^{-\alpha}$ as described by Eq. (4). The  amplitude  ratio  $C^+/C^-   =  n/2^{d/2}$, where $n$ is the number of order parameter components. Above $T_c$, phase and modulus fluctuations in the absence of the equilibrium value of the order parameter represent just two equivalent degrees of freedom of the scalar complex order parameter. Below $T_c$, the symmetry of the system decreases. The order parameter modulus fluctuations remain of the same diffusive type as above $T_c$, while the character of the phase fluctuations, in accordance with the Goldstone theorem, changes dramatically. It has been shown that the effect of phase fluctuations is dimension dependent. Accordingly, it has been demonstrated that the inclusion of phase fluctuations leads to a reduction in the degree of order in $d > 2$ and to its complete destruction in $d \leq 2$ \cite{landau, stanley, ma}. Carrying out the integral over the real and imaginary parts of the order parameter, one can find an expression for the fluctuation part of the free energy, which results in the disappearance of the temperature dependence of the phase fluctuation contribution and, calculating the second derivative, one can see that only the fluctuations of the order parameter modulus contribute to the specific heat. As a result the specific heat, calculated below $T_c$, turns out to be proportional to that found above. 

 The considerable success of MFT or SGA in conventional superconductors originates from the low value of the critical Ginzburg width ($\delta{t_G} \sim 10^{-14}$) \cite{amit}. Indeed, one observes that the specific heat does not present any divergence, but rather a compatible jump with the predictions of Eq. (4). Practically, critical phenomena should not be observable in usual conventional superconductors, which is not the case in HTS.
 
Due to the small values of coherence length and the high anisotropy (between the in-plane and out-of-plane coherence lengths $\xi_{\parallel}$, and $\xi_{\perp}$) $\gamma = \xi_{\parallel}/\xi_{\perp} = \sqrt{m_{\perp}/m_{\parallel}} \gg 1$, the temperature region of  critical  fluctuations in  the  cuprate  superconductors appears to be large and can be studied in detail.  For $d < 4$, the superconducting interaction in HTS has a short coherence length $\xi$ which allows conserving nonzero local Cooper pairs for $T  = T_c - \delta{T_c}$ ($\delta{T_c} > 0$), while $\xi \rightarrow \infty$ when $T \rightarrow T_c$, as demonstrated with models which take into account these local fluctuations of the Cooper pairs by construction \cite{ppapon}. Fundamentally, there are several mechanisms which can explain the reduction of the critical temperature in HTS, but the focal point of our concern is the corrections due to local fluctuations of the Cooper pairs (i.e. OPF) and their effects on the thermodynamic parameters.  Accordingly, it is convenient to discriminate corrections due to the OPF and write the result for the reduction/suppression of the transition temperature $\delta{T_c}$ in the form
\begin{equation}
\frac{\delta{T_c}}{T_c} = \upsilon_{dc},
\end{equation}
where $\upsilon_{dc}$ represents a dimensionless quantity which characterizes the width of the critical region about the transition temperature through the value of the scaled correction term at the transition point. Following the usual terminology, $n$=  1, 2, and 3 correspond, respectively, to the so-called \textquotedblleft Ising,\textquotedblright \textquotedblleft XY,\textquotedblright and \textquotedblleft Heisenberg\textquotedblright models. Contrary to the case of the basic GL model, these are discrete models defined on a lattice and are therefore suited for numerical simulations. The XY Hamiltonian describes phase transitions with a single complex order parameter where phase fluctuations dominate. This is opposite to the GL functional, which accounts for predominant weak fluctuations of the amplitude of the order parameter \cite{bok}. In 3D systems, fluctuations of the order parameter will, in general, vary in space. The GL free energy form can still be used, if we assume that fluctuations are small. In 1D- or 2D-systems, we will still take into account only amplitude fluctuations, with the assumption that the quadratic coefficient now takes into account the correlations of fluctuations and the dimension dependence of the order parameter. Taking into account the fact that superconducting interactions are not uniform in HTS and that the order parameter depends to a large extent on the geometry and the dimensionality (spatial dimension), we will assume an additional dimension dependence of the critical temperature $T_c(d)$, which, like the expressions of the critical exponents $\alpha$ and $\alpha'$ is an indication of the increased fluctuation effects as the dimensionality of the system decreases \cite{dikande}.  However, taking into account all fluctuation degrees of freedom involves a certain spatial averaging procedure that allows us to pass from the microscopic to mesoscopic and macroscopic levels, by adding a new quantity to the scaled quadratic coefficient. Then, the new scaled quadratic coefficient $\epsilon^{\ast}(d, T)$ of $F_{GL}$ replacing $\epsilon$ is defined as
\begin{equation}
\epsilon^{\ast}(d, T) = \epsilon + \upsilon(d, T),
\end{equation}
where the additional scaled coefficient $\upsilon(d, T)$, which includes a new contribution to the ($d$, $q$)-mode autocorrelation function, is determined by solving the following self-consistent equation \cite{roger}
\begin{equation}
\upsilon(d, T) = \mathcal{K}_d\bigg[\frac{T}{T_c}\bigg]\bigg(\epsilon + \upsilon(d,T) \bigg)^{\frac{d}{2} - 1}.
\end{equation}
The integral correction constant $\mathcal{K}_d$ related to stacking faults and other defects intrinsic to the layered structure of HTS is given by 
\begin{equation}  
\mathcal{K}_d \propto \xi_0^{-d}\int_0^{x_c}\frac{x^{d-1}}{1+x^2}\cdot dx. 
\end{equation}  
This quantity takes into account the fact that the magnitude and the temperature dependence of the fluctuation corrections strongly depend on the impurity concentration and the anisotropy of the electron spectrum \cite{gauzzi, anatoly, varlamov, varlamov2} and that superconducting properties show possibly related anomalies. The taking into account of the cutoff $x_c$ allows to regularize the model and is used to absorb infinities arising in the integral correction constant \cite{anatoly, varlamov2, kleinert2}. The present approach with cutoff included can be used to account for the deviations of experimental data on amorphous alloys  and low dimensional superconducting materials from the predictions of the standard GL theory \cite{gauzzi, bok}. Eq. (9) which presents a self-consistent condition with respect to the dimensionless correction term $\upsilon(d, T) \propto \langle  
|{\Psi}|^2\rangle_{d, T}$, is derived by making use of the renormalized Gaussian partition function and by taking into account the expected fact that it does not explicitly depend on the wave number $q$ (for more details, see especially the appendix in \cite{roger} and also the references therein). The correction quantity $\upsilon(d, T)$ is responsible for the cross-over from mean-field to critical behavior and corresponds to the variance of the order parameter at a single point in space evaluated at any temperature.  
 
It is well known that the transitions in two and three dimensions are radically different. The decrease of the critical temperature is specially pronounced in the case of bad metals such as synthetic metals and optimally doped or slightly underdoped HTS compounds \cite{emery, gdeutscher}. In order to determine the plausible influence of the OPF on the critical temperature, let us estimate the  static susceptibility of the system. From the thermodynamic definition, the inverse static susceptibility is given by the following analytical expression
\begin{equation}
{\chi^{\ast}}^{-1}=  a_0T_c[\epsilon + \upsilon(d, T)].
\end{equation}
Accordingly, the temperature at the critical point where the static susceptibility diverges is determined as
\begin{equation}
T^{\ast}_c = T_c(1 - \upsilon_{dc}).
\end{equation}
$\upsilon_{dc}  = \upsilon(d, T^{\ast}_c) $ characterizes the crossover from classical mean-field critical behavior to fluctuation-dominated critical behavior and provides an insight into the critical behavior (phase transition) of the typical system. Although the critical value $\upsilon_{dc}$ is useful for the evaluation of the Ginzburg criterion (which indicates in a semi-quantitative manner the range within which the distance from the modified Gaussian transition temperature is important \cite{amit}), a rigorous calculation which consists firstly in solving Eq. (9) according to the dimension, and secondly in inserting the obtained result in the equation: 
\begin{equation}
\epsilon + \upsilon(d, T) = 0,
\end{equation}
 allows us to derive the expresion of $T^{\ast}_c$ and the following results for $\upsilon_{dc}$:
\begin{eqnarray}
\left\{  \begin{array}{llll}
\upsilon_{1c} \sim 1, & \textrm{for}\phantom{...} d = 1 \\
\upsilon_{2c} = \frac{\mathcal{K}_{2D}}{1 + \mathcal{K}_{2D}}, & \textrm{for} \phantom{...} d = 2\\
\upsilon_{3c} \approx 0, & \textrm{for} \phantom{...} d = 3\\
\upsilon_{dc} = 0, & \textrm{for} \phantom{...} d \geq 4.\\
  \end{array}\right.
\end{eqnarray}
$\mathcal{K}_{2D} \propto \ln(1 + x^2_c)/\xi^{2}_0$ is the 2D integral correction constant. It is worth noting that in 1D and 3D cases, a numerical method is used to approximate the solutions. The superconducting observables generally change as a result of finite size effects, since the phase transition is smeared over some temperature range. The susceptibility of bulk materials is assumed to be infinite at the critical temperature. This will naturally never be observed in an experiment, although the signal can be huge. A limited size not only reduces the maximum, but also shifts it to a lower temperature. The density of the Cooper pairs is also influenced by finite sizes and it shows a rounding instead of an abrupt decrease at $T_c$, which makes it hard to determine the true critical temperature. The most striking consequence of limiting the physical extensions is seen in the critical temperature; merely decreasing the dimension can shift $T_c$ from several degrees to tens of Kelvin or suppress the superconducting order completely \cite{ppapon, ahlberg}. Obviously, Eq. (14) establishes the importance of OPF and shows that for $d \leq 3$, $T_c$ is a characteristic scale of temperature related to thermal fluctuations rather than the transition temperature. More precisely, for $d \leq 3$, Eqs. (12) and (14) show that the modified value of the critical temperature $T^{\ast}_c$ is shifted to lower temperatures compared to $T_c$. It appears also that for $d \geq 4$, $\upsilon_{dc} = 0$, which means that above dimension 4, the present SCA acts exactly like the MFT, whereas it predicts correctly the universal quantities for dimension 4 and below. In the case of 4D systems, the correction term $\upsilon(d, T)$ is reduced to a linear function of temperature as $\upsilon(d, T) \propto (T - T_c)$ and the renormalized quadratic coefficient becomes $a^{\ast}(T) = a^{\ast}_0T_c\epsilon$ with, however, $a^{\ast}_0 \neq a_0$. Analyzing the critical region, the critical width $\delta{T_c} =  T_c - T^{\ast}_c$ increases as the dimensionality of the system decreases. For two- and three- dimensional systems, the renormalized functional leads to two superconducting transitions, one at $T_c$ where the system becomes superconducting and locks to produce a state with very short-range phase coherence and the second at $T^{\ast}_c < T_c$, where the superconducting phase locks to produce a state with long-range phase coherence. Only the lower transition is a true phase transition, with a divergent correlation length \cite{beloborodov, raboutou, ebner, buckel}. In 3D systems, the scaled width $\upsilon_{3c} = \frac{\delta{T_c}}{T_c}$ is of the order of $10^{-14}$ to $10^{-10}$, much too small to be accessible experimentally. In such a case, the SCA acts like a quasi mean-field approximation and is a useful tool of estimating the importance of fluctuations in a given superconductor. However, there is a difference in the thermodynamic observables where the SCA-results differ quantitatively from the MF ones. As we shall see in the next section, intrinsic critical fluctuations in 3D lead in certain cases to a divergence of the specific heat jump at the critical point. This divergence, expected from the 3D-$XY$ model, is visible in the high-$T_c$ cuprates because of their short coherence length \cite{deutscher}. 

In the particular case of 2D lattice systems, while recovering critical fluctuations through the $\mathcal{K}_{2D}$-term, the modified critical temperature can be expressed as 
\begin{equation}
T^{\ast}_c = T_c(1 + \mathcal{K}_{2D})^{-1}. 
\end{equation}
\begin{figure}
\begin{center}
\includegraphics[scale=0.6]{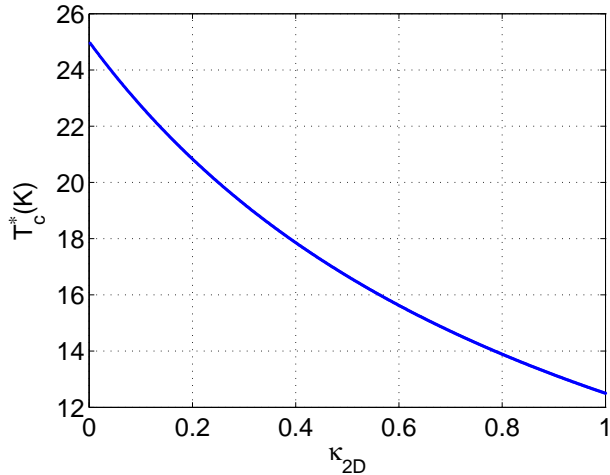}
\caption{The modified critical temperature $T^{\ast}_c$ plotted against the fluctuation-term $\mathcal{K}_{2D}$ for 2D lattice systems. The curve indicates the decrease of the renormalized critical temperature as intrinsic critical fluctuations of the system increase. It is assumed that $T_c$ = 25K as an indication.}
\end{center}
\end{figure}
 In the Ginzburg region, the fluctuation correction becomes important. We observe a factor of $1 \leq \tau  \leq 2$ decrease in $T_c$ and an order of magnitude increase in $\upsilon_{2c}$ between $\mathcal{K}_{2D} \approx 0$ and $\mathcal{K}_{2D} = 1$ (as shown in Fig. 1), resulting in a progression from relatively weak 2D fluctuations for $\mathcal{K}_{2D} \approx 0$ to strong 2D critical fluctuations over a wide temperature range. The cross-over from weak 2D to strong 2D critical behavior can be achieved with a fairly modest increase in $\mathcal{K}_{2D}$, which implies the existence of an intermediate high-dimensional regime between two-dimensional regimes \cite{gauzzi}. In the framework of weakly localized fluctuations, intrinsic critical fluctuations no longer affect the electron correlations, thus, $\upsilon_{2c} \rightarrow 0$ and $T^{\ast}_c \rightarrow T_c$. However, the small coherence lengths, high anisotropy, high transition temperatures, and quasi-2D nature of HTS greatly enlarge the temperature region in which OPF are important. A detailed analysis of critical fluctuations can provide important information regarding the dimensionality and the order parameter of the superconducting state. It is possible to obtain a complete suppression of the transition temperature due to large values of $\mathcal{K}_{2D}$. The superconducting phase transition in such a case is expected to belong to the 2D-XY model which does not exhibit any true long-range order, but does, however, undergo a Kosterlitz-Thouless transition at a finite temperature from a high-temperature phase where $\upsilon(d, T)$ has an exponential decay to a low-temperature phase with quasi-long-range order where $\upsilon(d, T)$ has a power-law decay \cite{roger, minnhagen, rice}. However, this behavior is not very probable for 2D-cuprate superconductors according to the GL model which accounts for predominant weak fluctuations of the amplitude of the order parameter. 

Of particular significance is the problem associated with the absence of phase transitions in a 1D system. It was well known that phase transitions involving a classical order parameter, such as the liquid-gas transition in a system of particles with finite-range interactions, cannot occur in 1D systems. Stimulated by a suggestion of the possibility of superconductivity in polymeric systems, it was shown by Rice \cite{rice} and more rigorously by Hohenberg \cite{hohenberg} that this could be extended to include the superconducting phase transition \cite{devreese}. In the framework of our model, the 1D-corrections are more pronounced and true long-range order from the breaking of a continuous symmetry is absent in accordance with the Mermin-Wagner-Hohenberg theorem. The modified static susceptibility $\chi^{\ast}(d = 1, T)$ becomes infinite at $T^{\ast}_c \sim 0$ K, showing that there is no phase transition when the spatial dimensionality is restricted to one dimension. Eq. (14) shows that $\upsilon_{1c} = \frac{\delta{T_c}}{T_c} \rightarrow 1$ for 1D systems. Mathematically, this circumstance implies that interactions between fluctuations are significant and that the quartic term and higher-order terms in $F_{GL}$ are no longer negligible. However, it is important to note that we take into account only OPF. This means that the ground state obtained is marginal and can be easily destroyed by any (other) small corrective term (interlayer coupling or axial anisotropy), which should be inherent to all real materials (such as cuprate superconductors).  Real systems are not isolated 1D systems but are composed of parallel chains that are in some sense coupled. Because of this such system can undergo phase transitions at finite temperature exhibiting 3D long-range order, although this will usually occur at a temperature significantly lower than $T_c$. Although, the quantity $\upsilon(d, T)$ is confined at the Gaussian level, its presence allows a low temperature behavior and it is appropriate for model calculations in a number of critical systems, especially those with relatively small size or with the presence of inhomogeneities, reducing the effective size of the coherent region \cite{tuszyinski, zinn-Justin}.

\section{Order parameter fluctuations and dimension effects on the specific heat (jump) of high-temperature superconductors}
\noindent

Studies of the fluctuation and dimension effects at phase transitions in the HTS provide valuable information concerning the mechanisms responsible for superconductivity and enable the derivation of certain constraints for the microscopic theory. Generally, according to the essential singularities exhibited by the thermodynamic quantities characteristic of phase transitions with in some cases a probable signature of global broken symmetry, there is a strong evidence of observed anomalies in the behavior of thermodynamic quantities in terms of microscopic models and interactions such as a loss of entropy $\Delta{S}^{\ast}_d$, a specific heat $\Delta{C}^{\ast}_d$ jump and a gain in renormalized free energy $\Delta{F}^{\ast}_d$ between the high and low temperature phases. Here, we are  only considering the electronic component, assuming that the ion dynamics is basically not affected by this  transition. From its expressions in the high and low temperature phases, the free energy difference is determined as
\begin{equation}
\Delta{F}^{\ast}_d = -a_0^2T^2_c\frac{[\epsilon + \upsilon(d, T)]^2}{2b_0}.
\end{equation}
By using Eq.  (16), the increase in the entropy associated with the suppression of the order parameter at $T^{\ast}_c$ can be obtained as follows:
\begin{equation}
\Delta{S}^{\ast}_d = \frac{a^2_0T_c}{b_0}\bigg(1 + T_c\frac{\partial{\upsilon(d, T)}}{\partial{T}} \bigg)\bigg(\epsilon + \upsilon(d, T) \bigg)\Bigg|_{T = T^{\ast}_{c}} = 0,
\end{equation}
showing that there is no change in the entropy at the modified transition point $T^{\ast}_c$. It appears clearly that the energy required to destroy the order of the system at the transition point or the latent heat is equal to zero. This result is not surprising, since for the second order phase transition, fluctuations suppress the transition, but conserve entropy. Therefore, the specific heat jump modified by fluctuations has the following expression
\begin{equation}
\Delta{C}^{\ast}_d = \frac{a^2_0T}{b_0}\Bigg[\bigg(1 + T_c\frac{\partial{\upsilon(d, T)}}{\partial{T}}\bigg)^2 + T^2_c\Big(\epsilon + \upsilon(d, T) \Big)\frac{\partial^2{\upsilon(d, T)}}{{\partial{T}}^2}\Bigg],
\end{equation}
which results from the integration over all fluctuation degrees of freedom.  Within the framework of the SCA, while taking into account the fact that $\Big(\epsilon + \upsilon(d, T)\Big)\Big|_{T = T^{\ast}_{c}} = 0$,  the anomalous part of the specific heat is given at $T^{\ast}_{c}$ by
\begin{equation}
\Delta{C^{\ast}_0} = \Bigg(1 - \upsilon_{dc} \Bigg)\Bigg(1 + T_c\frac{\partial\upsilon(d, T)}{\partial{T}}\Big|_{T = T^{\ast}_{c}} \Bigg)^2\Delta{C_0},
\end{equation}	
where $\Delta{C}_0$ is the mean-field specific heat jump defined in Eq. (6). For 1D and 2D systems, one gets
\begin{eqnarray}
 \left\{  \begin{array}{ll}\vspace{0,5cm}
\Delta{C}^{\ast}_0 \simeq 0,  & \textrm{for 1D systems \phantom{...} (a)}\\ \vspace{0,5cm}
\Delta{C}^{\ast}_0 = (1 + \mathcal{K}_{2D})\Delta{C}_0,  & \textrm{for 2D systems \phantom{...} (b)}.\\
  \end{array} \right.
\end{eqnarray}
 For 3D systems, the resolution of Eq. (9) leads to two solutions and we obtain at the critical temperature the following expression: 
\begin{eqnarray}
\Delta C^{\ast}_0 =  \left\{  \begin{array}{ll}\vspace{0,5cm}
0,  & \textrm{(a)}\\ \vspace{0,5cm}
4(1 + \mathcal{K}^2_{3D})^2\Delta{C}_0 & \textrm{(b)}.\\
  \end{array} \right.
\end{eqnarray}
Taking into account the contribution of fluctuations, it can be concluded that, if the transition takes place, the anomalous part of the specific heat will not be detected experimentally, particularly for 1D systems, and 3D systems in certain cases (see (a) of Eqs. (20) and (21)). The modified specific heat as T approaches $T^{\ast}_{c}$ from above is proportional to the MF specific heat jump, with a prefactor which depends on the dimension and microscopic parameters of the system. The solution (b) of Eq. (21) shows that the jump in the ratio of the modified and the MF specific heats at the transition point is at least 4.  This result is interesting because it leads to nontrivial predictions and is in reasonable agreement with certain experiments as will be discussed shortly. During the preparation of compounds like $\mathrm{YBCO}$, it is difficult to avoid the appearance of impurity phases. In some of them, such as the $\mathrm{Y}_2\mathrm{Cu}_2\mathrm{O}_5$ phase, the low-temperature specific heat is 10-100 times larger than the electronic specific heat in the $\mathrm{YBCO}$ compounds \cite{junodbook, plakida}. The present approach leads to an asymptotic behavior for temperatures appreciably far from $T_{c}$, while in the crossover region $\delta{T_c}$ or for finite size, the results obtained differ quantitatively from that of SGA or MFT. 

If the true test of the validity of an approach is in its comparison with experiments, another is the establishment and evaluation of the importance of the obtained results. Therefore, in order to proceed further with our investigation, we discuss the same issue for cuprates, whose transitions indeed apparently belong to the same universality class.  For these superconductors, we expect an even larger departure of the experiments from the above universal predictions because of the extremely short correlation length. For this reason, we demonstrate the usefulness of the results obtained in this paper through the cuprate superconductors, which present the case of the contribution of OPF to the electronic specific heat. Moreover, for the cuprate superconductors and in order to attain a phenomenological interpretation of the experimental results, we perform the renormalized Gaussian approximation where $F_{GL}$ is again truncated to second order resulting to the factorization of the partition function, and by integrating over all fluctuation degrees of freedom, the modified GL specific heat of the system is factorized as follows:

\begin{eqnarray}
&&C^{\ast}_{GL} =  \left\{  \begin{array}{llll}\vspace{0,5cm}
\frac{1}{2}\eta_d{(\xi^+_0)^{-d}}\Big[\frac{T}{T_c}\Big]^2\Big(1 + T_c\frac{\partial\upsilon(d, T)}{\partial T}\Big)^2\Big(\epsilon + \upsilon(d, T)\Big)^{-\alpha} \hspace{1.3cm} \textrm{for}\phantom{...} T > T^{\ast}_c. \\
\Delta C^{\ast}_0 +   \eta_d{(\xi^-_0)^{-d}} \Big[\frac{T}{T_c}\Big]^2\Big(1 + T_c\frac{\partial\upsilon(d, T)}{\partial T}\Big)^2\Big\vert\epsilon + \upsilon(d, T)\Big\vert^{-\alpha'} \hspace{0.5cm} \textrm{for} \phantom{...} T < T^{\ast}_c,\\
  \end{array}\right.
\end{eqnarray}
 $\Delta C^{\ast}_0  = g(\mathcal{K}_d)\Delta{C}_0 $ is the modified specific heat jump and $\eta_d$ is the same integral quantity defined in Eq. (4) for the Gaussian approximation and which is dimension-dependent. The modified specific heat jump is proportional to the MF one, with a prefactor
\begin{eqnarray}
g(\mathcal{K}_d) = \left\{  \begin{array}{lll}
0, & \textrm{for}\phantom{...} d = 1 \\
(1 + \mathcal{K}_{2D}), & \textrm{for} \phantom{...} d = 2 \\
4(1 + \mathcal{K}^2_{3D})^2, & \textrm{for} \phantom{...} d = 3, \\
  \end{array}\right.
\end{eqnarray}
which obviously leads to the enhancement,  suppression or rounding of the specific heat jump at the transition according to the dimension and the nature of interactions. The modified specific heat jump increases linearly for $d = 2$ and shows nearly exponential behavior for $d = 3$ (see Fig. 3). Now, the dominant behavior of $C^{\ast}_{GL}$ close to $T^{\ast}_c$ is obtained through the power law $C^{\ast}_{GL} \sim {\Big\vert \frac{T - T^{\ast}_c}{T_c} + \Big(\upsilon(d, T) - \upsilon_{dc} \Big)  \Big\vert}^{d\nu-2}$  giving rise to another temperature dependence of the specific heat and thermodynamic quantities in general. The correction terms
\begin{displaymath}
\Big(1 + T_c\frac{\partial\upsilon(d, T)}{\partial T}\Big)^2 \phantom{..} \textrm{and}  \phantom{..} \upsilon(d, T), 
\end{displaymath}
bring significant contributions to the specific heat.  Indeed, the quantity $\Big(1 + T_c\frac{\partial\upsilon(d, T)}{\partial T}\Big)^2$ indicates either a small or a large specific heat step as the case may be, as observed in certain superconductors \cite{sergey}.
 
  A different deviation from MF behavior is seen in the  transition of $\mathrm{YBCO}_{7-\delta}$ single crystal for $0 \leq \delta \leq 0.18$, as shown  in  Refs. \cite{bok, loram, meingast, junod, fisherfisher}. For example, large values of $\delta{T_c}$ and a strong 2D critical behavior are clearly defined as reported by Loram et al \cite{loram}. One also notes the sharp decrease of the specific heat jump at the transition which might be attributed to OPF. It is possible that some of this rounding may actually be due to the inhomogeneity of the sample as one goes from fully oxygenated, overdoped, to slightly underdoped samples, but it may also indicate the fluctuation-dominated critical regime \cite{bok}. The jump reported experimentally by Fisher et al \cite{fisherfisher} was found to be rounded, on the scale of order, a tenth of a degree K, for $\delta$ = 0. Due to the fact that the conducting $\mathrm{CuO}$ chains in $\mathrm{YBCO}_{7 - \delta}$ equally play a crucial role in enhancing the inter-planar coupling, relative changes in the phonon term due to $\delta$ in $\mathrm{YBCO}_{7-\delta}$ can be reliably determined and the fluctuation term clearly established with confidence over a wide temperature range \cite{loram}.
  
	Whereas investigations indicate a considerable weight of the specific jump in $\mathrm{YBCO}_{7-\delta}$, possibly with a modified temperature dependence, the jump is absent in the specific heat of the highly anisotropic $\mathrm{BSCCO}$ compound where the more radically different form of the specific heat transition given by the first solution (a) of Eq. (21) is well observed \cite{deutscher}. We note that both specific heat \cite{junod} and thermal expansion \cite{meingast} data find a nearly symmetric anomaly at the critical point with practically no jump component. This critical behavior which is better described by Bose condensation \cite{alexandrov} can also be well described by a GL theory renormalized by intrinsic critical fluctuations at the Gaussian level as discussed here. In $\mathrm{BSCCO}$ compounds, the  heat  capacity  transition  looks  very  much  different from that in $\mathrm{YBCO}_{7-\delta}$. In $\mathrm{YBCO}_{7-\delta}$, the ratio $\gamma$ is about 5, while in $\mathrm{BSCCO}$, it is about 30 \cite{deutscher}. The effective number of degrees of freedom is less significant in $\mathrm{BSCCO}$ than in $\mathrm{YBCO}_{7-\delta}$. As a result, the Cooper pairs in quasi-two dimensions fall off more rapidly than in three dimensions with increasing temperature around the critical point. This difference in behavior is also apparent in the variation of the critical temperature with the superfiuid density. In the $\mathrm{BSCCO}$ compound, there is no range of superfiuid density since it remains constant, contrary to what is seen in $\mathrm{YBCO}_{7-\delta}$ \cite{bok, annette, poole}.

 Another deviation from MF behavior predicted here concerns the ratio $\Delta{C_0}/T_c$ (between the specific heat jump and the critical temperature). With the standard GL theory, this ratio is a constant, while for high-$T_c$ superconductors it strongly depends on critical fluctuations. Basing our analysis on a renormalized Gaussian approach that takes into account both the influence of a nonuniform $T_c$ and the microscopic relation between the GL parameters (i.e. $\frac{a^2_0}{b_0} = \frac{8\pi^2}{7\zeta(3)}\nu$ where $\zeta(x)$ is the Riemann zeta-function), the modified ratio in 3D case (taking into account Eq. (21)) is given by 
\begin{equation}
\Delta{C^{\ast}_0}/T^{\ast}_c = \frac{32\pi^2}{7\zeta(3)}\nu(1 + \mathcal{K}^2_{3D})^2. 
\end{equation}
This result gives clear evidence for a systematic increase in the electronic specific heat jump with increasing superconducting fluctuations and describes a more detailed GL analysis yielding values for coherence lengths and critical temperatures. This remarkable behavior is well observed in a new class of transition metal oxide superconductors of the $\beta$-pyrochlore oxide type, such as $\mathrm{KOs}_2\mathrm{O}_6$ \cite{yonezawa, zenji}. The jump at the critical temperature is much larger than that in other superconductors \cite{kazakov}. It is possible that some of this enhancement may actually be due to a large lattice contribution, which survives down to low temperatures, and which enhances the $C/T$ ratio near the critical temperature, but it may also indicate the fluctuation-dominated critical regime and many anomalous features found in $\mathrm{KOs}_2\mathrm{O}_6$. These pyrochlores commonly posses characteristic Fermi surfaces involving a strong tendency for nesting \cite{kune}. To account for this nonlinear behavior of the specific heat, the taking into account of critical fluctuations is necessary and the Sommerfeld constant, which characterizes these materials should take at the critical point the following form: 
\begin{equation}
 \gamma(T^{\star}_c) = g(\mathcal{K}_{d})\gamma_0,
\end{equation}
where $\gamma_0 \propto \frac{a^2_0}{b_0}$ is the Sommerfeld constant extrapolated to zero temperature from the superconducting-state. $\mathcal{K}_d$ determines the critical line and hence incorporates both the asymptotic critical behavior and the crossover to the regular regime. 
\begin{figure}
\begin{center}
\includegraphics[scale=0.6]{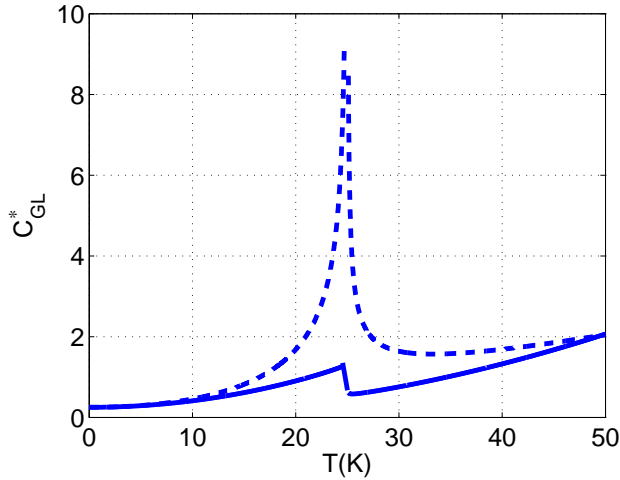}
\caption{Specific heat as a function of the temperature for 3D-lattice systems. The dashed curve represents the specific heat with MF critical exponent $\vert\alpha\vert = 0.5$ while the solid line represents the specific heat with 3D-$XY$ critical exponent $\vert\alpha\vert = 0.013$ according to Eq. (22).}
\end{center}
\end{figure} 

In general, depending on the coherence length, different types of critical behavior at a superconducting phase transition can be distinguished. Here, we are concerned with the short-coherence length cuprates, which, combined with their small density of Cooper pairs has raised the question of the existence of critical fluctuations in the whole fluctuation region, and consequently, the breakdown of the MF regime \cite{bok}. In the case of purely Gaussian fluctuations, no divergence of the mean-field jump is expected to occur, whereas (as appears to be the case here), the predictions suggest a divergence of the fluctuation contribution with critical exponent $\alpha = 2 - d/2 = 0.5$ for $d = 3$ (see the dashed line in Fig. 2). In the case of critical fluctuations described by the 3D-$XY$ model, a logarithmic divergence of the jump is expected. Experimental data on $\mathrm{YBCO}_{7-\delta}$ single crystals reported in Refs. \cite{pasler, overend} are in better agreement with the latter prediction than with the first one, as discussed in Ref. \cite{junod}. Also, measurements of the specific heat in $\mathrm{YBCO}_{7-\delta}$ compounds, with sharp transitions, suggest a crossover of the critical exponent from the 3D-mean-field value ($\vert\alpha\vert = 0.5$) to the value $\vert\alpha\vert \leq 0.018$ \cite{pasler} given by the 3D-$XY$ model. The zero-field specific heat of $\mathrm{YBCO}_{7-\delta}$ is well described by the 3D-$XY$ model, with the same critical parameters as those found for liquid $^4\mathrm{He}$, over a temperature range of 10 K above and below   $T_c$ \cite{overend}. This range agrees well with the range over which the electronic specific heat has been claimed to vary logarithmically with  temperature. 

 Although confined at the Gaussian level, the quantity $\upsilon(d, T)$ represents a deviation from the standard Gaussian model, as it contains the recipe for taking into account the quartic and higher-order terms in the case of strongly localized fluctuations. The taking of account of higher-order terms can result in an adjustment of critical exponents in the vicinity of the critical point. However, the change of value of the critical exponents does not occur \textquotedblleft brutally\textquotedblright at a \textquotedblleft given temperature\textquotedblright, it occurs within a certain interval of temperatures close to the true critical point $T^{\ast}_c$ (and often far from $T_c$). Using the 3D-$XY$ critical exponent $\vert\alpha\vert = 0.013$, we perform practically the same fits as in $\mathrm{YBCO}_{7-\delta}$ compounds \cite{bok, deutscher} (see the solid line of Fig. 2) with the difference however, that the specific heat jump now takes  into account the intrinsic critical fluctuations.  For 3D-systems, as displayed in Fig. 3, the jump in the ratio of the modified and the mean-field specific heats rapidly increases with the fluctuation-term, showing nearly exponential behavior.  The crossover from no divergence to a divergence of the mean-field jump can be achieved with a fairly modest increase in $\mathcal{K}_{d} = \mathcal{K}_{3D}$ expected from high-Tc cuprates because of their short coherence length. In such a case, the dashed and solid curves in Fig. 2 can merge, thus rendering the renormalized GL and the 3D-$XY$ models indistinguishable. In such a case, it is difficult to discriminate definitively if such a divergence is really associated with the value of the critical exponent throughout the critical region or if it is due to intrinsic effects associated with sample inhomogeneities, which may be present even in apparently very good compounds. This principal aspect will be discussed in more details elsewhere.
\begin{figure}
\begin{center}
\includegraphics[scale=0.5]{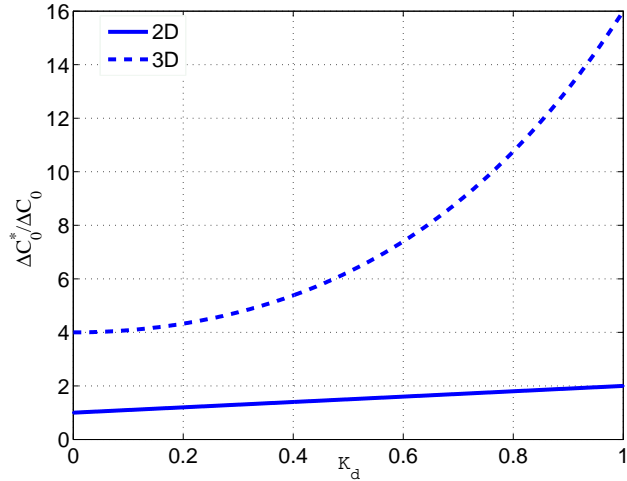}
\caption{The scaled specific heat jump ($\Delta{C^{\ast}_0}/\Delta{C_0}$) as a function of the fluctuation term $\mathcal{K}_{d}$ for 2D and 3D  lattice systems. The dashed curve represents 3D lattice systems while the solid curve shows the corresponding curve for 2D lattice systems.}
\end{center}
\end{figure} 

In the case of 2D-systems, the modified specific heat jump increases in a rather slow manner with increasing $\mathcal{K}_{d} = \mathcal{K}_{2D}$. The solid line plotted in Fig. 3 shows a weak fluctuation-dependent behavior at low $\mathcal{K}_{2D}$. It can be seen that the dependence of the modified specific heat jump on critical fluctuations is linear and a logarithmic divergence of the jump is expected according to $\mathcal{K}_{2D}$. Amongst other things, of central importance is the rounding behavior of the specific heat jump at the transition for short-coherence length systems. This behavior can be understood in terms of OPF. The GL model in the free-field approximation and the 3D-XY model are two classic examples of universality class with one complex order parameter. These two classes are suitable for describing the weak and strong fluctuation regimes respectively. However, with a modified quadratic coefficient, the renormalized GL model can also describe strong fluctuation regimes as described in this paper.

\section{Conclusion}
\noindent

The dimension and temperature dependence of the specific heat is evaluated using a renormalized method of calculation of averages, which retains a significant part of the spatial features of the fluctuations of the Cooper pairs. The obtained expressions of this specific heat are discussed as the sample dimensions are lowered from three dimensional bulk materials through two and one dimensions. The approach presented confirms the fact that the standard GL functional is no longer a valid approximation for high-temperature superconductors. This is because the thermal fluctuations of the Cooper pairs near the critical point are large due to the short coherence length, which follows directly from the high critical temperature through Heisenberg's uncertainty principle. From the developed self-consistent approach, the prefactor of the logarithmic increase of the specific heat jump is determined for 2D systems. For 3D systems, the renormalized specific heat jump is at least four times higher than the mean-field specific heat jump as shown in Fig. (3). We believe that this behavior is also due to a large lattice contribution, which survives down to low temperatures and which enhances the specific jump near the critical point as observed in the $\beta$-pyrochlore oxide $\mathrm{KOs}_2\mathrm{O}_6$. The case where the specific heat jump obtained is zero as observed in $\mathrm{Bi}_2\mathrm{Sr}_2\mathrm{CaCu}_2\mathrm{O}_{8 + \delta}$ suggests that if the transition takes place, the anomalous part of the specific heat can not be detected experimentally. Another interesting feature of the calculations is the existence of a microscopic parameter $\upsilon_{dc}$, which incorporates both the asymptotic critical behavior and the crossover to the regular regime. The present model reproduces the MFT for dimension four and above with thermodynamic observables modified by intrinsic critical fluctuations. It also correctly predicts the universal quantities for dimensions lower than four, taking into account the Mermin-Wagner-Hohenberg theorem. At all temperatures below and above $T_c$, the SCA matches the superconducting observables, thus providing a unified picture for both the Gaussian and the critical (non-Gaussian) regimes. However, we recall that our fluctuation studies probe only the integral of the renormalized Gaussian spectrum taking into account the expected fact that the newly introduced non-universal quantity $\upsilon(d, T)$ which represents a deviation from the standard Gaussian model does not explicitly depend on the wave number $q$. This quantity does not modify the critical exponents but affects, however, the characteristic length scales, the specific heat jump and the crossover region where critical exponents are supposed to vary. It would therefore be essential to determine more precisely the momentum dependence of the renormalized Gaussian spectrum and compare it with the Lorentzian dependence of the conventional MF spectrum. This should enable us to determine other corrective terms of the conventional GL Hamiltonian which are expected to give a better description of the superconducting state of cuprates. In this respect, a systematic study of other fluctuation properties, such as the fluctuation diamagnetism, would be important. Fluctuation studies on other short-coherence length superconductors would also be useful for exploring possible new aspects of the physics of phase transitions in short-coherence length systems.

\begin{center}
\textbf{acknowledgements}
\end{center}

This work is partially supported by the Abdus Salam International Centre for Theoretical Physics (ICTP, Trieste, Italy), through the Office of External Activity (OEA)-Prj - 15. The ICMPA is also in partnership with the Daniel Iagolnitzer Foundation (DIF), France. We wish to acknowledge many stimulating discussions with Professor Emeritus Samuel Domngang, Chairman of the Cameroon Academy of Sciences.

\end{document}